\documentclass[11pt]{article}
\usepackage{moriond,epsfig}

\def\Journal#1#2#3#4{{#1} {\bf #2}, #3 (#4)}

\def\PRD{{\em Phys. Rev.} D}

\def\be{\begin{equation}}
\def\ee{\end{equation}}
\def\bea{\begin{eqnarray}}
\def\eea{\end{eqnarray}}

\begin{document}
\vspace*{4cm}
\title{Higgs production cross-section in a Standard Model with four generations at the LHC }

\author{Charalampos Anastasiou$^1$, Stephan Buehler$^1$, Elisabetta Furlan$^2$ , 
  Franz Herzog$^1$, Achilleas Lazopoulos$^1$}

\address{$^1$Institute of Theoretical Physics, ETH Zurich, 8093 Zurich, Switzerland\\
$^2$Physics Department, Brookhaven National Laboratory, Upton, NY 11973, USA}

\maketitle 
\abstracts{
We present theoretical predictions for the Higgs boson production cross-section
via gluon fusion at the LHC in a Standard Model with four generations.  
We include  QCD corrections through NLO retaining the full dependence 
on the quark masses, and the NNLO corrections in  the
heavy quark effective theory approximation. We also include  
electroweak corrections through three loops. 
Electroweak and bottom-quark contributions 
are suppressed in comparison to
the Standard Model with three generations.
}

\section{Introduction}

The Tevatron and the LHC are becoming increasingly  sensitive to
potential signals of a Standard Model (SM) Higgs boson. The highest  
sensitivity  is achieved for a Higgs  boson of  mass  
\mbox{$m_h \sim 2 \, m_W$}, 
where the branching ratio of the Higgs decaying into  a pair of $W$ bosons 
is close to one.  

The most significant production  channel for these 
searches  is gluon fusion. This is a loop-mediated  process, and in
the Standard Model the corresponding amplitudes are dominated by 
top-quark loops.  Gluon fusion may be sensitive to new coloured, heavy
particles 
with a large coupling to the Higgs
boson.  Limits on the Higgs production cross-section from ongoing
experimental searches are model-specific, and a dedicated  theoretical 
prediction to Standard Model extensions is often required. 

A simple extension of the Standard Model may include  a fourth family
of quarks and leptons.  This model is theoretically constrained to not
very large fourth-generation quark masses, in order to preserve
perturbativity~\cite{Marciano:1989ns}.  
To evade limits from precision electroweak
tests, the mass difference of the fourth-generation quarks  is also
restricted to be small~\cite{Kribs:2007nz}. Finally, the fourth-generation neutrino
is required to have a mass greater than half  the mass of the Z-boson
from the LEP limits on the
invisible Z-boson decay width. 

One might find little theoretical appeal in a naive Standard Model with
four generations. As in the Standard Model, it  suffers from the
hierarchy problem, and in addition it  introduces a rather awkwardly 
heavy neutrino. Nevertheless, its predictions  are  rather
spectacular in modifying the Higgs boson cross-section at hadron
colliders and can be tested easily with Tevatron and early LHC data. 
The Tevatron  has  published limits on the Higgs  boson
cross-section in this model, excluding a wide range of Higgs boson
masses~\cite{Aaltonen:2010sv}. 
Recently the CMS collaboration carried out a similar study~\cite{Chatrchyan:2011tz}.  

A precise calculation of the Higgs boson gluon fusion
cross-section has been made in Ref.~\cite{Anastasiou:2010bt}, 
where numerical results 
have been given for the Tevatron. In this work, we compute the
cross-section at the LHC with center of mass energy $\sqrt{s} = 7\,
{\rm TeV}$. 

\section{Calculation details}

In the Standard Model, gluon fusion Higgs production is mediated  by massive
quarks and electroweak gauge bosons.  
The  dominant contribution is given by top-quark loops. Bottom-quark loops
and  $W,Z$ loops yield a small contribution, and quarks lighter 
than the bottom  can  be neglected.

The gluon fusion cross-section through massive quark loops 
has been computed  through NLO in perturbative QCD 
both in the heavy quark effective theory~\cite{Dawson:1990zj,Djouadi:1991tka} 
and retaining the full top-mass dependence~\cite{Graudenz:1992pv,Spira:1995rr}. 
The NLO corrections are
large, motivating the calculation of the cross-section through NNLO
in perturbative QCD~\cite{Harlander:2002wh,Anastasiou:2002yz,Ravindran:2003um}. 
The NNLO corrections have first been
computed in the heavy quark effective theory (HQET). Recently,
subleading corrections in the inverse top-quark mass
expansion have been calculated~\cite{Harlander:2009mq,Pak:2009dg}, 
validating the quality of the HQET
approximation. Other  corrections include the  exact two-loop 
electroweak contributions~\cite{Actis:2008ug}  and mixed QCD  and electroweak 
corrections~\cite{Anastasiou:2008tj}  using an effective theory approach. 

 A  precise prediction of the  Higgs  boson cross-section in fixed  order
perturbation theory may be obtained  by combining  the  above
contributions, with an estimated uncertainty of about 
$\pm 10\%$ due  to missing higher order  effects. Calculations
resumming factorizable contributions of  infrared nature  are in
agreement with this  error uncertainty~\cite{Catani:2003zt,deFlorian:2009hc}. 
A similar  uncertainty at the
LHC  and  slightly larger at the Tevatron is   due to parton
distribution functions.

In this work, we compute the cross-section in a Standard Model with
four fermion generations.  
The following points make this calculation 
different from the calculation in the 
Standard Model. 
\begin{itemize}
\item  The gluon fusion amplitudes  receive  contributions from the
  up and down quarks of the fourth generation. We compute these
  contributions  through NLO retaining  the exact mass
  dependence. In the  strict HQET  limit, the cross-section is increased 
  by a factor of nine with respect to the SM cross-section.
\item  Bottom-quark contributions are suppressed with respect to the
  SM by roughly  a factor of three. We  compute  bottom-quark 
  contributions  exactly  through NLO in perturbative  QCD.     
\item Through NLO, the Wilson coefficient  of the HQET  operator is
  three  times  the Standard Model Wilson coefficient. At NNLO, the
  Wilson coefficient depends on the masses of the the heavy quarks
  and it  also receives  contributions from three-loop diagrams 
  containing two different heavy quarks~\cite{Anastasiou:2010bt}.
\item Two-loop electroweak corrections due to light quarks are
 given by the same diagrams as in the Standard Model~\cite{Aglietti:2004nj}. They
 predominate the  
 electroweak corrections for a light Higgs  boson.  
 They are suppressed  in the  Standard Model with four
 generations compared to the three-generation SM. In the 
 SM, additional electroweak corrections  due
 to the top quark are very small for a light Higgs boson
 and have been computed  in Ref.~\cite{Actis:2008ug}. 
Here  we  include  the full two-loop Standard Model electroweak
corrections of Ref.~\cite{Actis:2008ug}, but we  neglect the two-loop electroweak 
corrections  with quarks of the fourth generation.
\end{itemize}
We compute the NNLO corrections in the effective theory
approximation. We normalize  these corrections to the Born cross-section with the exact
mass  dependence on the masses of the top-quark and the quarks  of
the fourth generation. We do not include  bottom-quark loops at NNLO. 

\section{Predictions  for the gluon fusion cross-section 
for $\sqrt{s} = 7 \, {\rm TeV}$}
In this section we present  our  numerical results for the gluon
fusion cross-section in a  Standard Model with four generations  at  
the current  LHC  energy of $7\, {\rm TeV}$ and for Higgs  boson masses in
the range  $m_h \in \left[ 110\,{\rm GeV}, 300\,{\rm GeV} \right]$. 
The parameter space of the Standard Model with four generations  is
quite limited. The model becomes non-perturbative  for values of
the  heavy quark masses roughly above $\sim 500 \, {\rm GeV}$~\cite{Marciano:1989ns}.  A large mass
splitting  of the quarks of the fourth generation induces large
corrections  to oblique electroweak parameters~\cite{Kribs:2007nz}.  To comply
with these constraints  we present  the cross-section for two
scenarios, 
where the mass of the fourth-generation down quark is chosen as 
\begin{equation}
\mbox{scenario 1: } m_{d_4} = 300 \,  {\rm GeV} \; , \qquad 
\mbox{scenario 2: } m_{d_4} = 400 \,  {\rm GeV} \;.
\end{equation}
The mass of the fourth-generation up quark given by~\cite{Kribs:2007nz}
\begin{equation}
m_{u_4} - m_{d_4} = 50 \, {\rm GeV} + 10 \, {\rm GeV} \, \times \,
\log\left( 
\frac{m_h}{115 \, {\rm GeV}}
\right) \, {\rm GeV} \;. 
\end{equation}
For the top- and bottom-quark masses we take
\begin{equation}
m_t = 172 \, {\rm GeV} \; ,  \qquad m_b(m_b) = 4.2 \, {\rm GeV} \; .
\end{equation}
We use the MSTW08 NNLO parton densitities~\cite{Martin:2009iq} and quote the
``$\alpha_s$+pdf'' uncertainty at the $90\%$ confidence level (CL). 
To estimate the uncertainty due to higher order perturbative effects, 
we vary the renormalization and factorization scales in the range 
$\frac{m_h}{4} \leq \mu_r=\mu_f  \leq m_h$. The scale variation and pdf
uncertainty are  very similar to the  equivalent  uncertainties in the 
Standard  Model.  We present  our results in  Table~\ref{tab:cross-sections}. 
\begin{table}
\begin{center}
\begin{tabular}{|c|c|c|c|c|c|c|}
\hline
{\rm GeV} 
&  
$
\begin{array}{c}
\sigma[pb] \\ 
\mbox{scenario 1}
\end{array}
$
&
$
\begin{array}{c}
\sigma[pb] \\ 
\mbox{scenario 2}
\end{array}
$
& 
{\tiny 
$
\begin{array}{c}
\delta ^{(+)} ({\rm pdf}+\alpha_s) \\
{\rm MSTW08} \\
90\% \, {\rm CL}
\end{array}
$
}\%
&  
{\tiny 
$
\begin{array}{c}
\delta ^{(-)} ({\rm pdf}+\alpha_s) \\
{\rm MSTW08} \\ 
90\% \, {\rm CL}
\end{array}
$
}\%
& $\delta ^{(+)} ({\mu})$\%
& $\delta ^{(-)} (\mu)$\% \\ 
\hline
105 & 202.33 & 201.39 & 7.9 & -7.6 & 9.2 & -9.7 \\ 
110 & 183.41 & 182.51 & 7.9 & -7.6 & 9.0 & -9.7 \\ 
115 & 166.85 & 165.97 & 7.9 & -7.6 & 8.9 & -9.6 \\ 
120 & 152.27 & 151.41 & 7.9 & -7.6 & 8.7 & -9.6 \\ 
125 & 139.38 & 138.54 & 7.9 & -7.6 & 8.6 & -9.6 \\ 
130 & 127.93 & 127.12 & 7.9 & -7.6 & 8.5 & -9.5 \\ 
135 & 117.72 & 116.93 & 7.9 & -7.6 & 8.4 & -9.5 \\ 
140 & 108.59 & 107.81 & 7.9 & -7.6 & 8.3 & -9.5 \\ 
145 & 100.39 & 99.628 & 7.9 & -7.6 & 8.2 & -9.4 \\ 
150 & 93.002 & 92.253 & 7.9 & -7.6 & 8.1 & -9.4 \\ 
155 & 86.298 & 85.563 & 7.9 & -7.6 & 8.0 & -9.4 \\ 
160 & 80.091 & 79.371 & 7.9 & -7.6 & 7.9 & -9.4 \\ 
165 & 74.221 & 73.516 & 7.9 & -7.7 & 7.8 & -9.4 \\ 
170 & 68.920 & 68.228 & 8.0 & -7.7 & 7.8 & -9.3 \\ 
175 & 64.249 & 63.570 & 8.0 & -7.7 & 7.7 & -9.3 \\ 
180 & 60.000 & 59.333 & 8.0 & -7.7 & 7.6 & -9.3 \\ 
185 & 56.080 & 55.424 & 8.0 & -7.8 & 7.6 & -9.3 \\ 
190 & 52.493 & 51.849 & 8.1 & -7.8 & 7.5 & -9.3 \\ 
195 & 49.246 & 48.612 & 8.1 & -7.8 & 7.4 & -9.3 \\ 
200 & 46.306 & 45.681 & 8.1 & -7.9 & 7.4 & -9.2 \\ 
205 & 43.620 & 43.005 & 8.1 & -7.9 & 7.3 & -9.2 \\ 
210 & 41.153 & 40.546 & 8.2 & -7.9 & 7.3 & -9.2 \\ 
215 & 38.878 & 38.279 & 8.2 & -8.0 & 7.2 & -9.2 \\ 
220 & 36.776 & 36.185 & 8.2 & -8.0 & 7.2 & -9.2 \\ 
225 & 34.832 & 34.249 & 8.3 & -8.0 & 7.2 & -9.2 \\ 
230 & 33.031 & 32.454 & 8.3 & -8.1 & 7.1 & -9.2 \\ 
235 & 31.357 & 30.788 & 8.3 & -8.1 & 7.1 & -9.2 \\ 
240 & 29.805 & 29.242 & 8.3 & -8.1 & 7.0 & -9.2 \\ 
245 & 28.357 & 27.800 & 8.4 & -8.2 & 7.0 & -9.1 \\ 
250 & 27.009 & 26.459 & 8.4 & -8.2 & 7.0 & -9.1 \\ 
255 & 25.751 & 25.206 & 8.5 & -8.3 & 6.9 & -9.1 \\ 
260 & 24.577 & 24.038 & 8.5 & -8.3 & 6.9 & -9.1 \\ 
265 & 23.480 & 22.945 & 8.5 & -8.3 & 6.9 & -9.1 \\ 
270 & 22.455 & 21.926 & 8.6 & -8.4 & 6.8 & -9.1 \\ 
275 & 21.495 & 20.970 & 8.6 & -8.4 & 6.8 & -9.1 \\ 
280 & 20.598 & 20.078 & 8.6 & -8.4 & 6.8 & -9.1 \\ 
285 & 19.756 & 19.241 & 8.7 & -8.5 & 6.7 & -9.1 \\ 
290 & 18.969 & 18.457 & 8.7 & -8.6 & 6.7 & -9.1 \\ 
295 & 18.232 & 17.725 & 8.8 & -8.6 & 6.7 & -9.1 \\ 
300 & 17.541 & 17.037 & 8.8 & -8.6 & 6.7 & -9.1 \\
\hline        
\end{tabular}
\end{center} 
\caption{Gluon fusion cross-section in a
  Standard Model with four fermion generations. 
  The masses of the fourth generation quarks are
  chosen according to ``scenario 1''
  and ``scenario 2''. All cross-sections are computed with $0.1\%$ Monte-Carlo
  integration error or better.}
\label{tab:cross-sections}
\end{table} 

\section{Sensitivity to parton distributions}

Besides  the MSTW08 pdf set, other two NNLO parton distribution sets are currently made available 
by the GJR~\cite{JimenezDelgado:2009tv} and ABKM~\cite{Alekhin:2009ni} collaborations.   
In Table~\ref{tab:pdfs} we present the  central value of the cross-section for $\mu =
\frac{m_h}{2}$,  the scale variation uncertainty and the
pdf uncertainty (at the $68\%$ CL) for $m_h=110,165,200, 300\, {\rm GeV}$
and the ABKM09, GJR and MSTW08 pdf sets.  

\begin{table}[th]
\begin{center}
\begin{tabular}{|c||c|c|c|c|}
\hline
$\sigma [pb$]  &  
$\mbox{ABKM09}$& $\mbox{GJR}$ & $\left. \mbox{MSTW08}\right|_{68\%
  {\rm CL}}$ 
& $\left. \mbox{MSTW08}\right|_{90\%
  {\rm CL}}$  
\\
\hline
$m_h= 110\, {\rm GeV}$ & $167.59 \pm 3.0\%_{\rm pdf}$
                                     & $162.78 \pm 3.6\%_{\rm pdf}$ 
                                     & $183.41  \begin{array}{c} +4.0
                                       \\  - 3.1 \end{array}\%_{\rm
                                       pdf}$  
                                     & $ \begin{array}{c} +7.9  \\  - 7.6 \end{array}\%_{\rm pdf}$ \\ 
$m_h= 165\, {\rm GeV}$ & $66.130 \pm 3.3\%_{\rm pdf}$ 
                                     & $67.713 \pm 3.3\%_{\rm pdf}$ 
                                     & $74.221 \begin{array}{c} +4.0
                                       \\  - 3.3 \end{array}\%_{\rm
                                       pdf}$  
                                     & $ \begin{array}{c} +7.9  \\  - 7.7 \end{array}\%_{\rm pdf}$  \\
$m_h= 200\, {\rm GeV}$ & $40.634 \pm 3.6\%_{\rm pdf}$ 
                                     & $42.867 \pm 3.5\%_{\rm pdf}$ 
                                     & $46.306 \begin{array}{c} +4.1  \\  - 3.4 \end{array}\%_{\rm pdf}$ 
                                     & $ \begin{array}{c} +8.1  \\  - 7.9 \end{array}\%_{\rm pdf}$ \\
$m_h= 300\, {\rm GeV}$ & $14.768 \pm 4.7\%_{\rm pdf}$ 
                                     & $16.786 \pm 5.0\%_{\rm pdf}$  
                                     & $17.541 
                                    \begin{array}{c} +4.3  \\  - 3.9 \end{array}
                                        \%_{\rm pdf}$  
                                     & $ 
                                    \begin{array}{c} +8.8  \\  - 8.6 \end{array}
                                        \%_{\rm pdf}$ 
                                         \\
\hline
\end{tabular}
\end{center}
\caption{A  comparison for the gluon fusion cross-section in the 
``scenario 1'' of the four-generation Standard Model with
  the three available NNLO pdf sets: ABKM09, GJR and MSTW08.}
\label{tab:pdfs}
\end{table}
We  find  that the GJR and ABKM09 pfds give central values for the
cross-section which can be  up to $12\%$ smaller than the one of 
MSTW08. These differences  are larger than what anticipated from the  
quoted parton density uncertainties at the $68\%$ confidence level.
The MSTW08 provides  uncertainties  at the $90\%$ confidence level,
which yield  about twice as large an uncertainty for  the gluon fusion
cross-section. These  overlap, albeit marginally for lower Higgs
boson masses, with the uncertainties  of GJR and ABKM09.    

All three groups provide consistent determinations  of the parton 
distributions with self-consistent choices and assumptions, and 
there is no ``bullet-proof''  argument to choose one  over  the
others.  Nevertheless, we prefer  MSTW08 as our default  pdf. 
Our main reasons for this choice are that MSTW08 includes jet data directly sensitive 
to the gluon density and that their central value of $\alpha_s(m_Z)$ is 
in very good agreement with the world average \cite{Bethke:2009jm} 
and determinations from jet data at $e^+e^-$ colliders using NNLO jet cross-sections~\cite{Dissertori:2009qa}.
We also find  it  prudent to estimate the pdf uncertainty of the cross-section  
using parton density uncertainties at the   $90\%$ CL.  
The extraction of parton distribution functions will soon be assisted with early LHC data, 
which we hope will help to resolve the discrepances among the various pdf sets.

\section{Branching ratios}

For a complete  prediction of a Higgs  signal cross-section at
colliders the branching ratios of the Higgs boson decays to
observable final states are needed. Branching ratios  are significantly  modified
with respect  to the Standard Model when adding  a  fourth quark and
lepton generation.  

In this model, the partial decay width of the
Higgs boson to  gluons  is  enhanced  by a large factor which reaches
nine in the HQET limit.  This width dominates for a light Higgs boson.  
The decay widths to $WW$ and  $ZZ$ are significant  for  a  range of
values of the Higgs boson mass above ~$140 \,{\rm GeV}$. 
However, the corresponding branching ratios  are smaller than in the Standard
Model. These  decays  are  important, given that the Tevatron
and the LHC are  quite sensitive  and their  experimental 
study may  lead to a Higgs boson discovery with a modest  amount 
of data. In the four-generation SM, novel decays of a heavy Higgs boson 
to the leptons 
of the fourth generation emerge and assume  a  significant width. 

A systematic study of the Higgs boson decays has been made in
Ref.~\cite{Kribs:2007nz}, where the branching ratios have been 
computed by modifying the program {\tt HDECAY}~\cite{Djouadi:1997yw}. Tabulated 
results  from Ref.~\cite{Kribs:2007nz} for the branching ratios  of
the  Higgs  boson, corresponding to ``scenario 1'' and ``scenario 2'' 
of our study, can be found in Ref.~\cite{Aaltonen:2010sv}.   

\section{Conclusions}

In this work we  have studied the Higgs boson cross-section at the
LHC, in a Standard Model with a  fourth generation.  We have  computed 
the cross-section through NNLO in perturbative  QCD, including finite
quark-mass effects  and electroweak corrections  through NLO. We have
provided an estimate of the uncertainty due to higher order perturbative
effects, and have studied the sensitivity of the cross-section on
various NNLO parameterizations of the parton densities and their
uncertainties. Our results are of direct relevance to the ongoing
studies for  the discovery of the Higgs  boson at the LHC.

\subsection*{Acknowledgments}
Research supported by the  Swiss National Foundation under 
contract SNF 200020-126632 and the DOE under Grant DE-AC02-98CH10886.

\end{document}